\documentclass[11pt,osajnl2,preprint,showpacs,superscriptaddress]{article}
\usepackage{graphicx}
\usepackage{osajnl2}
\usepackage{color}
\usepackage[normalem]{ulem}
\begin{document}
\title{Two-beam nonlinear Kerr effect to stabilize laser frequency with sub-Doppler resolution}
\author{W. Soares Martins, H. L. D. de S. Cavalcante, T. Passerat de Silans$^*$, M. Ori\'{a}, M. Chevrollier}
\address{Laborat\'{o}rio de Espectroscopia \'{O}tica, DF-CCEN, Cx. Postal 5086 - Universidade Federal da Para\'{i}ba, 58051 - 900 Jo\~{a}o Pessoa - PB, BRAZIL}
\email{thierry@otica.ufpb.br}
\begin{abstract}
Avoiding laser frequency drifts is a key issue in many atomic physics experiments. Several techniques have been developed to lock the laser frequency using sub-Doppler dispersive atomic lineshapes as error signals in a feedback loop. We propose here a two-beam technique that uses non-linear properties of an atomic vapor around sharp resonances to produce sub-Doppler dispersive-like lineshapes that can be used as error signals. Our simple and robust technique has the advantage of not needing either modulation or magnetic fields.
\end{abstract}


\maketitle

\section{Introduction}

Many important atomic transitions, such as the D2 line of alkali atoms, have a linewidth of typically a few MHz and most monomode lasers emitting around one of these transitions have a narrower linewidth, thus meeting high resolution spectroscopy requirements.  However, in order to resonantly excite an optical transition ($\sim10^{14}$ Hz) it is necessary to tune the laser frequency with a relative precision of $\sim10^{-8}$. Resonant matching may not hold for a long time due to thermal and other small fluctuations of the laser parameters that result in frequency drifts and make the laser depart from resonance. Therefore, active control of the center frequency of the laser emission may be necessary, particularly in long-run experiments \cite{Banerjee04}.\\

Many different schemes to lock the laser frequency have been developed. The basic idea behind those techniques is to produce a dispersive-like error signal used to actively control the laser frequency. It can be interesting to generate dispersive sub-Doppler signals centered at hyperfine transitions to have an absolute frequency reference. Most of the techniques employed involve modulation with homodyne detection to generate dispersive lineshapes from the derivative of absorptive sub-Doppler lines, such as saturation absorption spectroscopy with frequency modulation \cite{Rovera94} and two-photon absorption in the presence of ac magnetic field \cite{Baluschev00}. On the other hand, the use of modulation can limit the feedback linewidth \cite{Yoshikawa03} and should be avoided for some purposes \cite{FM}. Techniques without modulations have been developed and use, for instance, the induced birefringence in the vapor \cite{Yoshikawa03} or the difference of absorption of Zeemam sub-levels in the presence of a magnetic field, the so-called sub-Doppler DAVLL technique \cite{Harris08,Tino}.\\

Recently, we have proposed to use nonlinear optical properties of alkali vapors to produce Doppler-broadened dispersive-like lineshapes by exploring the intensity dependence of the vapor´s refractive index due to a third-order effect in the field amplitude. We have called this technique ANGELLS \cite{Angells}, for Atomic Nonlinearly GEnerated Laser Lock Signal. Alkali vapors are very interesting nonlinear systems since the third-order susceptibility ($\chi^{(3)}$) can be easily modified by changing the field amplitude or by finely tuning the laser frequency across the resonance \cite{Bjorkholm74}. The dependence of $\chi^{(3)}$ with the detuning is the key element of the ANGELLS technique we have developed to lock the laser frequency with Doppler resolution \cite{Angells}. Moreover, the $\chi^{(3)}$ term can be drastically modified by the introduction of a second laser beam permitting sub-Doppler resolution. In this manuscript we report on a modified ANGELLS technique to produce sharp dispersive resonances, with homogeneous linewidth, in the intensity-dependent refractive index and use such a signal to lock the laser frequency to any hyperfine transition of the Rb D2 line. \\

\section{Exploring the intensity-dependent refractive index to obtain a dispersive lineshape}

\begin{figure}[p]
	\centering
		\includegraphics[width=0.95\linewidth]{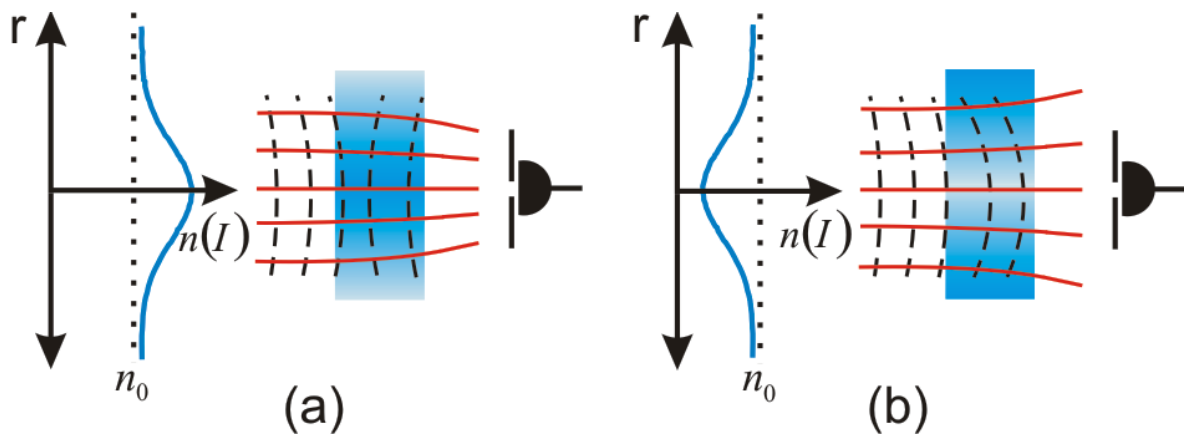}
	\caption{(Color online) Illustration of the effect of positive (a) and negative (b) increment in the nonlinear refractive index. Positive increment (a) results in self-focusing of a Gaussian beam, while negative increment (b) induces self-defocusing. Figure1.eps.}
	\label{Fig1}
\end{figure}

In the presence of strong laser beams an atomic vapor behaves as a Kerr medium, whose intensity-dependent refractive index is given by:
\begin{equation}
n=n_0+n_2I,
\end{equation}
where $I$ is the beam intensity. If one sends through the vapor a beam with Gaussian intensity profile:
\begin{equation}
I(r)=I_0e^{-\alpha r^2},
\end{equation}
with $r$ the beam radial coordinate, one ``prints'' in the vapor a radial index variation with the same spatial profile as the beam:
\begin{equation}
n(r)=n_0+n_2I_0e^{-\alpha r^2},
\end{equation}
which changes the beam shape in a process named self-lensing \cite{Ackemann98}. In other words, the laser changes the vapor index and the medium acts back to the radiation as a lens for the beam itself. For a positive $n_2$ the wave front travels more slowly in the center than in the edge, inducing the focusing of the beam (see Figure \ref{Fig1}a). Conversely, for a negative $n_2$, the wave front travels faster in the center, thus inducing defocusing (see Figure \ref{Fig1}b). The self-focusing/defocusing leads to a change in the power transmitted through an aperture placed past the atomic vapor cell, allowing this phase effect to be directly measured by a photodetector. \\


The nonlinear coefficient of the refractive index is a third order effect on the field amplitude and can be calculated as \cite{Boyd}:
\begin{equation}
n_2=\frac{3}{4n^2_0\epsilon_0c}\Re{(\chi^{(3)})}
\end{equation}
where $\epsilon_0$ is the vacuum permittivity, $c$ is the light speed and $\Re{(\chi^{(3)})}$ is the real part of $\chi^{(3)}$. The signal of $\Re{(\chi^{(3)})}$ changes across the atomic resonance, giving rise to a sign change of $n_2$ and of the self-lensing effect, which passes from self-defocusing (red detuned) to self-focusing (blue detuned). Detection of the laser beam after an aperture past a hot resonant atomic vapor thus results in a Doppler-broadened dispersive lineshape, used in the ANGELLS technique to lock the laser frequency \cite{Angells}.\\

\begin{figure}[p]
	\centering
		\includegraphics[width=0.95\linewidth]{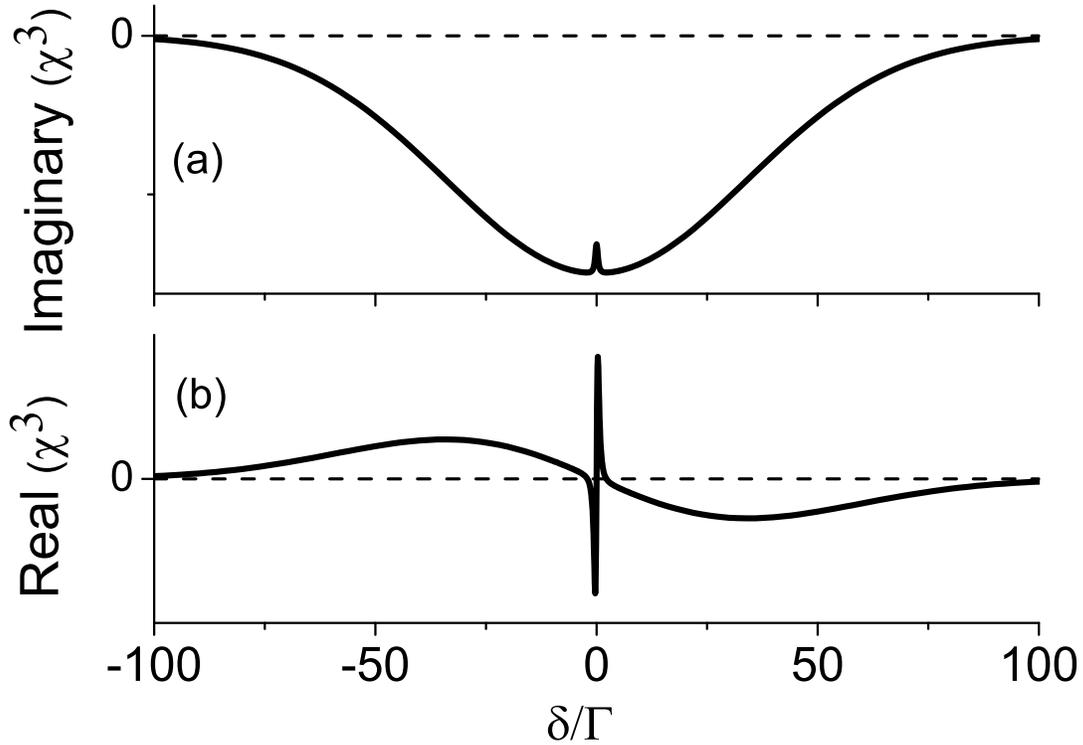}
	\caption{Imaginary (a) and real (b) parts of $\chi^{(3)}$ for a closed two-level system as a function of frequency detuning ($\delta$) normalized by natural linewidth ($\Gamma$). The lineshapes were calculated by numerical integration over a Maxwell-Boltzmann velocity distribution corresponding to a temperature of $70^\circ$C and with pump intensity $0.02I_S$ and probe intensity $0.1I_S$ ($I_S$ is the saturation intensity). Figure2.eps.}
	\label{FigT}
\end{figure}

Differently from \cite{Angells}, a second counter-propagating pump beam is introduced to produce a sharp dispersive lineshape. It velocity-selectively saturates the vapor and modifies the medium seen by the detected probe beam. We use the density matrix formalism for a closed two-level system to treat the interaction of the probe beam with this pump-modified medium. The $\chi^{(3)}$ is obtained from the medium polarization expression \cite{Boyd}:
\begin{equation}
P=N\left(\mu_{eg}\rho_{ge}+\mu_{ge}\rho_{eg}\right)=\epsilon_0\left(\chi^{(1)}+3\chi^{(3)}\left|E_P\right|^2\right)E_P \label{pol}
\end{equation}
where $\mu_{eg}$ is the electric dipole moment of the transition, $\rho_{eg}$ is the coherence term of the density matrix and $E_P$ is the probe beam electric field. Note that we are only interested in the term $\propto E_P^3$ in the third order susceptibility (eq. \ref{pol}), because it is the term that is responsible for the self-lensing effect of the probe beam. The medium modification induced by the pump beam is implicit in the $\chi^{(3)}$ value. The coherence term can be obtained from the steady state regime ($\dot{\rho}=0$) of the equation of motion for the density matrix \cite{Gea95}. For a hot atomic vapor, the Doppler effect must be taken into account by shifting the probe and the pump frequencies in opposite directions for a given velocity class and integrating over the Maxwell-Boltzmann velocity distribution. We have numerically solved and integrated the steady-state solutions for a closed two-level system, and we show in Figure \ref{FigT} the obtained real and imaginary parts of $\chi^{(3)}$. 

The imaginary part of $\chi^{(3)}$ is plotted in Figure \ref{FigT}a, where we see, superposed to the Doppler-broadened signal, a peak with homogeneous width corresponding to the saturation of atoms travelling perpendicular to the laser beams. The real and imaginary parts of $\chi^{(3)}$ are related by Kramers-Kronig relations \cite{Boyd}, therefore, a sharp saturation effect must also appear in the real part of the susceptibility, as a dispersive lineshape (see Figure \ref{FigT}b).

For a real system, the signal lineshape should include the hyperfine structure of the ground and excited levels. For the D2 transition of Rb vapor used in this work, the expected real part of the $\chi^{(3)}$ lineshape is a Doppler-broadened dispersive signal superposed to sharp dispersive lineshapes at the frequencies of the hyperfine and crossover resonances. Therefore, the change in the power transmitted through an aperture past the vapor and induced by those sub-Doppler resonances can be used to lock the laser to the corresponding frequencies within homogeneous width \cite{width}.

\section{Experimental set-up}

\subsection{Optical set-up}

\begin{figure}[p]
	\centering
		\includegraphics[width=0.95\linewidth]{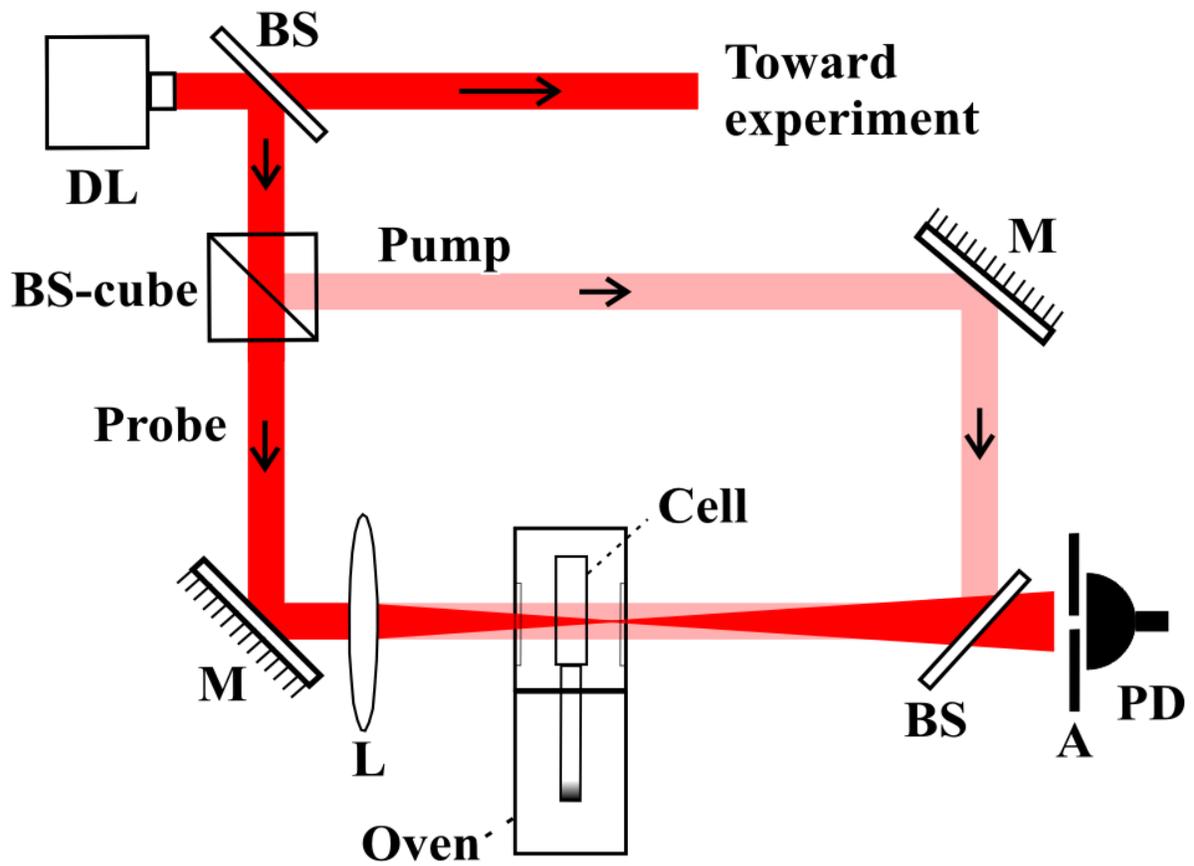}
	\caption{(Color online) Experimental set-up to produce sub-Doppler dispersive lineshapes. For the sake of clarity, the optical isolator and the auxiliary saturated absorption experiment are not shown. M are mirrors, BS are beamsplitters, L is the focusing lens, A is an aperture and PD is a photodetector. Figure3.eps.}
	\label{Fig2}
\end{figure}

The experimental set-up used to generate the saturated dispersive signal is illustrated in Figure \ref{Fig2}. The light source is a temperature- and current-stabilized, external-cavity diode laser. The semiconductor chip has an anti-reflection coating on its front side and the 15-mm external cavity is closed by a blazed diffraction grating with 1800 lines/mm. This laser system is tunable around the D2 line of rubidium atoms at 780 nm. A small fraction ($\sim1-2$ mW) of the laser power is deflected by a beamsplitter and used in the stabilization setup. A 60 dB optical isolator is placed at the laser exit in order to avoid optical feedback in the laser direction, from the counter-propagating beam and residual reflections. A cube beamsplitter (50/50) splits the beam into the probe and the pump beams, each one with $\sim 0.5-1.0$ mW. The probe beam with $\sim 1$-mm diameter goes through a lens ($\sim 100$ mm focal length), through the atomic vapor cell, which is placed $\sim20$ mm before the focal point, and through an aperture of diameter $\sim 2$ mm which allows around $20\%$ of the beam power to be transmitted and detected by an amplified photodiode. The focusing of the probe beam in the vapor allows us both to enhance the non-linear effects and to diminish the beam absorption by saturating the vapor. Just before the aperture, a beamsplitter reflects 90\% of the pump beam toward the vapor cell in the direction opposite to the probe beam. The pump beam is not focused so as to ensure a better overlap with the probe beam. A 10 mm-long cell containing a natural mixture of $^{87}$Rb and $^{85}$Rb is placed in an oven that allows us to control the atomic density via the temperature of the Rb reservoir ($T=70 \pm 1^\circ$C). The parameters focal length, aperture transmission, cell length, optical power, etc., given above are typical values, used to produce the results shown in this article, and are not optimized. We were able to generate clear sub-Doppler dispersive signals for a variety of set-up parameters around these typical values, the technique showing nice flexibility. \\

In order to provide a diagnostic signal, an auxiliary saturated absorption experiment was set up (not displayed in Figure \ref{Fig2}) and used as a frequency discriminator for analyzing the laser emission characteristics and the locking performance.

\subsection{Electronic feedback}

We have used in this experiment an external-cavity diode laser whose emission frequency can be finely controlled by applying a voltage to a piezo-electric (PZT) ceramic that controls the diffraction grating angle. In order to lock the laser frequency, a voltage is fed back to the PZT using, as an error signal, the dispersive lineshape detected by the photodiode.\\

The scheme of the electronic circuit that produces the electronic feedback is detailed elsewhere \cite{Angells} and will be described here in a simplified way: the photodetector signal has a dispersive-like shape with a non-zero voltage average value, corresponding to the off-resonance aperture transmission. We obtain an error signal centered at zero by subtracting a controlled reference voltage from the photodetector signal. This procedure allows us to finely choose the locking frequency inside the narrow homogeneous line by adjusting the reference voltage. This error signal is fed into a home-made electronic circuit having adjustable proportional and integral gains, whose output is sent to the PZT.

\section{Results}

\begin{figure}[p]
	\centering
		\includegraphics[width=0.95\linewidth]{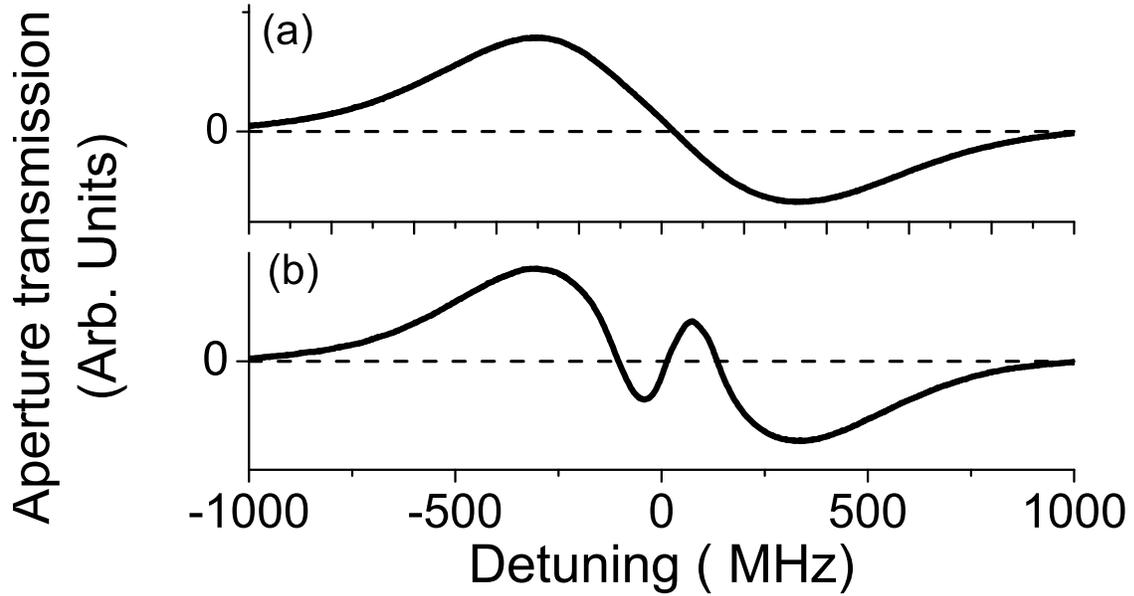}
	\caption{Transmission of the probe beam through the aperture without (a) and with (b) the counter-propagating pump beam when the frequency is scanned around the $^{85}$Rb $5S_{1/2}(F=2)\rightarrow 5P_{3/2}(F=1,2,3)$(unresolved) transitions. The probe and pump powers used are both $0.3$ mW, and the temperature of the Rb reservoir is at $T=70^\circ$C (atomic density of $2\times10^{12}$ at/cm$^3$). The zero line corresponds to the off-resonance transmission. Figure4.eps.}
	\label{Fig3}
\end{figure}

By scanning the laser frequency one can access the different hyperfine transitions from the hyperfine ground state levels of both $^{87}$Rb and $^{85}$Rb isotopes. In Figure \ref{Fig3} we show the transmission of the probe beam through the aperture when the laser frequency is scanned around the $^{85}$Rb $5S_{1/2}(F=2)\rightarrow 5P_{3/2}(F=1,2,3)$ transitions. Without the pump beam, a Doppler-broadened dispersive lineshape is observed as in \cite{Angells} (Figure \ref{Fig3}a). When the pump is on (Figure \ref{Fig3}b), a clear sub-Doppler dispersive lineshape appears, corresponding to saturation of velocity classes almost perpendicular to the beams. One can use this sub-Doppler resonance as an error signal to lock the laser frequency. Note that for this specific hyperfine transition, the lineshape is well described by our closed two-level model since the excited levels are not resolved due to collisional and power broadening.\\

\begin{figure}[p]
	\centering
		\includegraphics[width=0.85\linewidth]{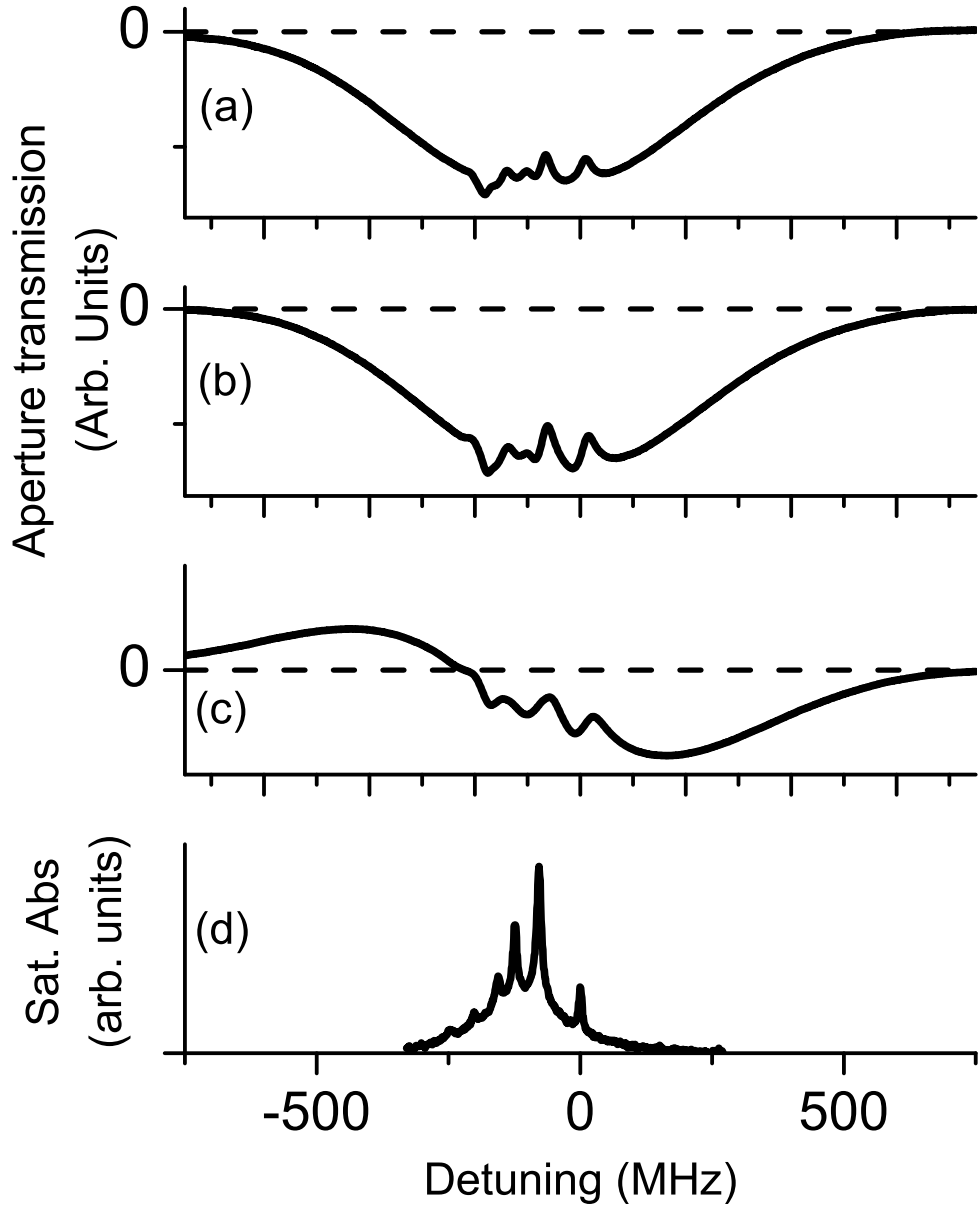}
	\caption{Transmission of the probe beam through the aperture when the laser frequency is scanned around the $^{87}$Rb $5S_{1/2}(F=1)\rightarrow 5P_{3/2}(F=0,1,2)$ transitions, for different probe beam powers and for a Rb vapor at $T=70^\circ$C (atomic density $\sim$ $2\times10^{12}$ at/cm$^3$): for a probe power of (a) 20 $\mu$W; (b) 60 $\mu$W; (c) 300 $\mu$W. (d) Homodyne detection of an amplitude-modulated saturated absorption reference signal. The zero line in (a),(b) and (c) corresponds to the off resonance transmission. Figure5.eps.}
	\label{Fig4}
\end{figure}

In Figure \ref{Fig4}, the signal corresponding to the $^{87}$Rb $5S_{1/2}(F=1)\rightarrow 5P_{3/2}(F=0,1,2)$ transitions is shown for different probe powers, together with a simultaneously-recorded reference saturated absorption spectrum (Figure \ref{Fig4}d). For a low probe power (Figure \ref{Fig4}a), the nonlinear contribution to the refractive index is negligible and the probe beam transmission detected past the aperture is only affected by first order absorption (imaginary part of $\chi^{(1)}$), whose sub-Doppler well resolved structures are due to saturation by the pump beam of selected velocity classes, corresponding thus to usual saturated absorption. For higher probe powers the product $n_2I$ becomes important and the signal is a mixture of absorption and dispersion (see Figure \ref{Fig4}b), exhibiting asymmetric lineshapes. To get a more symmetric signal one can either use a thinner vapor cell or increase the power of the probe beam, to enhance nonlinear dispersive effects against linear absorptive ones.  For still higher probe powers one observes well resolved dispersive-like sub-Doppler lineshapes at hyperfine transitions and crossover resonances frequencies, corresponding to saturation of selected velocity classes. We use those sharp dispersive lineshapes to lock the laser frequency.\\

\begin{figure}[p]
	\centering
		\includegraphics[width=0.95\linewidth]{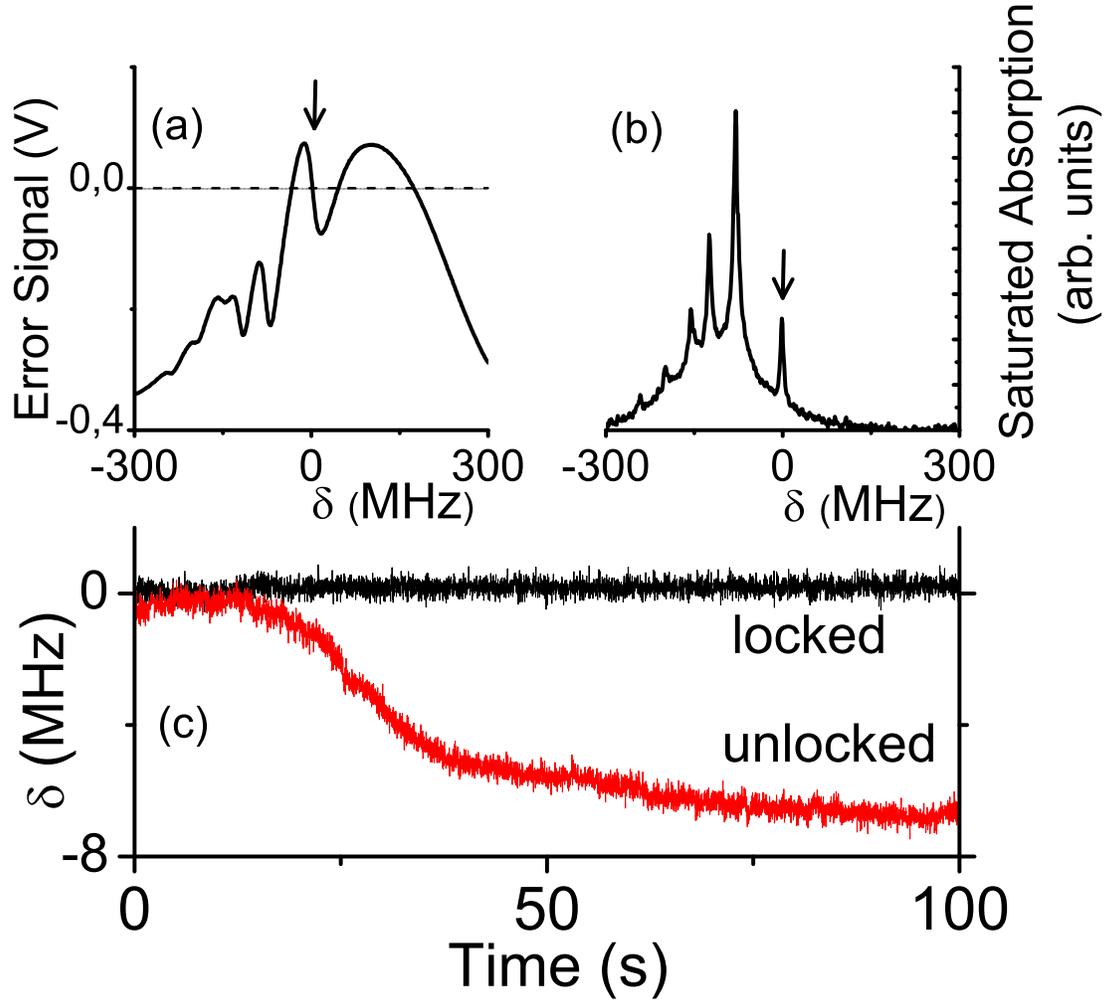}
	\caption{(Color online) (a) Error signal as a function of frequency detuning relative to $^{87}$Rb $5S_{1/2}(F=1)\rightarrow 5P_{3/2}(F'=2)$ hyperfine transition; (b) Saturated absorption from a reference cell; (c) Saturated absorption signal for the laser locked to the $^{87}$Rb $5S_{1/2}(F=1)\rightarrow 5P_{3/2}(F'=2)$ hyperfine transition as well as for an unlocked laser. The locking frequency is indicated by arrows in (a) and (b). Figure (c) shows locking times up to 100s to compare with the unlocked laser but locking times of more than one hour were reached with similar performance. Figure6.eps.}
	\label{Fig5}
\end{figure}

We produce the error signal by subtracting the adjustable reference voltage from the detector signal, in order to have a zero at the center of the hyperfine or crossover sub-Doppler dispersive lineshape at which the laser frequency will be locked. As an example, in Figure \ref{Fig5}a, the dispersive sub-Doppler signal corresponding to the hyperfine transition $5S_{1/2}(F=1)\rightarrow 5P_{3/2}(F=2)$ of $^{87}$Rb is brought close to zero and the laser frequency is then locked to this transition. The saturated absorption signal used as a reference to analyze the locking efficiency is shown in Figure \ref{Fig5}b.  In Figure \ref{Fig5}c the slow frequency drift is shown for the laser locked to this hyperfine transition as well as for the unlocked laser. One observes that the locked frequency remains at the desired position for time scales of minutes with a resolution corresponding to the laser linewidth of $\sim 1$ MHz, while for the unlocked laser the frequency departs from the homogeneous linewidth within a few seconds. We have locked the laser frequency during a typical run time of a few hours with the above resolution and have obtained similar performance for the different hyperfine and crossover transitions. Moreover, by modifying the setup parameters (laser power, cell temperature,...) it is possible to subtly make the homogeneous-width dispersive curve asymmetric in such a way that the center of the lineshape is at the flank of a particular hyperfine transition and allows us to lock the laser at a frequency slightly detuned from the resonance, for instance at the optimal $\Gamma/2$ detuning used to cool atomic vapors \cite{MOT}.\\

\section{Conclusion}
We have described a technique to lock the frequency of a diode laser to a hyperfine transition of the Rb D2 line. The technique uses nonlinear properties of an atomic vapor to produce sub-Doppler dispersive lineshapes in a two-beam extension of the ANGELLS technique developed before \cite{Angells}. The saturation of selected velocity classes results in homogeneous-width dispersive signals at hyperfine and crossovers transitions. Such narrow signals allow us to lock the laser to hyperfine and crossover transitions frequencies, without the need of a second reference cell to monitor the atomic resonances. The proposed technique has very simple implementation and is robust in the sense that it has little sensitivity to parameters fluctuations (room temperature and humidity, beam alignment fluctuations, etc.), the laser frequency remaining locked for more than one hour. The technique has simpler implementation than well known ones like saturated absorption spectroscopy \cite{Rovera94} or DAVLL \cite{Harris08,Tino}, with the clear advantage of not requiring modulation techniques or magnetic fields to be performed, at the expense of only a small power broadening of the sub-Doppler lines.\\

ACKNOWLEDGEMENT: This work was partially funded by Conselho Nacional de Desenvolvimento Cient\'{i}fico e Tecnol\'{o}gico (CNPq, contract 472353/2009-8), Coordena\c{c}\~{a}o de Aperfei\c{c}oamento de Pessoal de N\'{i}vel Superior (CAPES/Pr\'{o}-equipamentos) and FINEP. H.L.D.S.C., W.S.M and M.C. thank Brazilian agency CNPq for financial support.

\newpage
LIST OF FIGURES CAPTIONS:\\

Fig. 1. (Color online) Illustration of the effect of positive (a) and negative (b) increment in the nonlinear refractive index. Positive increment (a) results in self-focusing of a Gaussian beam, while negative increment (b) induces self-defocusing. Figure1.eps.\\

Fig. 2. Imaginary (a) and real (b) parts of $\chi^{(3)}$ for a closed two-level system as a function of frequency detuning ($\delta$) normalized by natural linewidth ($\Gamma$). The lineshapes were calculated by numerical integration over a Maxwell-Boltzmann velocity distribution corresponding to a temperature of $70^o$C and with pump intensity $0.02I_s$ and probe intensity $0.1I_S$ ($I_S$ is the saturation intensity). Figure2.eps.\\

Fig. 3. (Color online) Experimental set-up to produce sub-Doppler dispersive lineshapes. For the sake of clarity, the optical isolator and the auxiliary saturated absorption experiment are not shown. M are mirrors, BS are beamsplitters, L is the focusing lens, A is an aperture and PD is a photodetector. Figure3.eps.\\

Fig. 4. Transmission of the probe beam through the aperture without (a) and with (b) the counter-propagating pump beam when the frequency is scanned around the $^{85}$Rb $5S_{1/2}(F=2)\rightarrow 5P_{3/2}(F=1,2,3)$(unresolved) transitions. The probe and pump powers used are both $0.3$ mW, and the temperature of the Rb reservoir is at $T=70^\circ$C (atomic density of $2\times10^{12}$ at/cm$^3$). The zero line corresponds to the off-resonance transmission. Figure4.eps.\\

Fig. 5. Transmission of the probe beam through the aperture when the laser frequency is scanned around the $^{87}$Rb $5S_{1/2}(F=1)\rightarrow 5P_{3/2}(F=0,1,2)$ transitions, for different probe beam powers and for a Rb vapor at $T=70^\circ$C (atomic density $\sim$ $2\times10^{12}$ at/cm$^3$): for a probe power of (a) 20 $\mu$W; (b) 60 $\mu$W; (c) 300 $\mu$W. (d) Homodyne detection of an amplitude-modulated saturated absorption reference signal. The zero line in (a),(b) and (c) corresponds to the off resonance transmission. Figure5.eps.\\

Fig. 6. (Color online)  (a) Error signal as a function of frequency detuning relative to $^{87}$Rb $5S_{1/2}(F=1)\rightarrow 5P_{3/2}(F'=2)$ hyperfine transition; (b) Saturated absorption from a reference cell; (c) Saturated absorption signal for the laser locked to the $^{87}$Rb $5S_{1/2}(F=1)\rightarrow 5P_{3/2}(F'=2)$ hyperfine transition as well as for an unlocked laser. The locking frequency is indicated by arrows in (a) and (b). Figure (c) shows locking times up to 100s to compare with the unlocked laser but locking times of more than one hour were reached with similar performance. Figure6.eps.

\clearpage

\clearpage
\newpage

\clearpage
\newpage

\clearpage
\newpage

\clearpage
\newpage

\clearpage
\newpage


\begin{thebibliography}{99}
\bibitem{FM} Fast external phase modulation, as, for example, using electro-optic modulators, can solve the problem of limited feedback bandwidth, but with the drawback of losing simplicity and making the experiment more expensive.
\bibitem{width} As the technique described is induced by nonlinear effects, thus requiring strong fields, the sub-Doppler signal might be power-broadened. The technique resolution is thus larger than the natural (or pressure-broadened) width.
\end{thebibliography}
\end{document}